\documentclass[twoside]{ilcws10}
\usepackage[latin1]{inputenc}
\usepackage[dvips]{graphicx,epsfig,color}
\usepackage{wrapfig,rotating}
\usepackage{amssymb,amsmath,array}

\newcommand{\comment}[1]{}
\def\be{\begin{equation}}
\def\ee{\end{equation}}
\def\beq{\begin{equation}}
\def\eeq{\end{equation}}
\def\bea{\begin{eqnarray}}
\def\eea{\end{eqnarray}}

\def\agr{{\mathrm Re} a_\gamma}
\def\azr{{\mathrm Re} \Delta a_Z}
\def\agi{{\mathrm Im} a_\gamma}
\def\azi{{\mathrm Im} \Delta a_Z}
\def\bgr{{\mathrm Re} b_\gamma}
\def\bzr{{\mathrm Re} b_Z}
\def\bgi{{\mathrm Im} b_\gamma}
\def\bzi{{\mathrm Im} b_Z}
\def\bgtr{{\mathrm Re} \tilde b_\gamma}
\def\bztr{{\mathrm Re} \tilde b_Z}
\def\bgti{{\mathrm Im} \tilde b_\gamma}
\def\bzti{{\mathrm Im} \tilde b_Z}
\def\pl{P_L}
\def\plbar{\overline P_L}

\def\peff{P^{\rm eff}_L}

\def\plm{p_3}
\def\plp{p_4}

\begin{document}

\title{Probing anomalous $ZZH$ and $\gamma ZH$ interactions at an $e^+e^-$ linear collider using polarized beams}
 \author{Saurabh D. Rindani, Pankaj Sharma\vspace{.3cm}\\ Theoretical Physics Division, Physical Research Laboratory,\\
 Navrangpura, Ahmedabad - 380 009, India.}
% \author{Pankaj Sharma} \email{pankajs@prl.res.in}

% \affiliation{Theoretical Physics Division, Physical Research Laboratory,
% Navrangpura, Ahmedabad - 380 009, India.}

\maketitle

\begin{abstract}
 We examine anomalous $ZZH$ and $\gamma ZH$ interactions in the process $e^+e^-\to HZ$ 
followed by $Z\to l^+l^-$ at a linear collider. 
We study the effects of beam polarization, both longitudinal and transverse,
 in probing these anomalous couplings. We 
study various correlations constructed out of the initial and final state 
lepton momenta  
and the spins of electron and positron. We evaluate 
the sensitivity of these observables 
to probe anomalous couplings at the linear collider.
 \end{abstract}
\section{Introduction}\vspace{-0.3cm}
Although the standard model (SM) provides us an almost complete theory of electro-weak interactions, 
its mechanism of spontaneous breakdown of symmetry is still not well understood. Higgs, 
the center pillar of SM, has not been discovered yet. It is most likely that the Large Hadron Collider (LHC) 
will discover the SM Higgs. However, many alternative scenarios beyond SM offer solutions to 
electroweak symmetry breaking (EWSB) by incorporating more than one Higgs boson. Therefore, 
it is essential to carry out  a thorough study of the properties of the Higgs, which is extremely 
difficult task at LHC. However, these precision tests can be meticulously performed at 
an $e^+e^-$ linear collider. The International Linear Collider (ILC), which is at the design stage 
and likely to become a reality \cite{LC_SOU}, is definitely the well suited for this purpose.
% \begin{wrapfigure}{r}{0.6\columnwidth}
% \vspace{-0.4cm}
%  \begin{center}
%  \includegraphics[scale=0.5]{diagram1.eps}
% \end{center}
% \vspace{-0.8cm}
% \caption{Fig. (a) and (b) represent $e^+e^-\rightarrow ZH$ process with SM couplings and anomalous $VZH$ couplings.}
% \vspace{-0.4cm}
% \end{wrapfigure}

% \vspace{-0.3cm}
% % \section{Structure of anomalous VZH couplings}
% % \vspace{-0.4cm}
 At the lowest order in SM the $ZZH$ vertex is simply point-like whereas the 
$\gamma ZH$ vertex vanishes. Interactions beyond SM can modify this vertex 
by means of a momentum-dependent form factor, as well as by adding more 
complicated momentum-dependent forms of anomalous interactions considered 
in \cite{zerwas,hagiwara,Skjold}. Demanding Lorentz invariance, 
the most general coupling structure of $VZH$ vertex may be expressed as:
\begin{align}
 \begin{split}
  \Gamma_{\mu\nu}=g_Vm_Z\Big[a_Vg_{\mu\nu}+\frac{b_V}{m_Z^2}(k_{1\nu}k_{2\mu}-g_{\mu\nu}k_1\cdot k_2)
+\frac{\tilde{b}_V}{m_Z^2}\epsilon_{\mu\nu\alpha\beta}k_1^{\alpha}k_2^{\beta}\Big]
 \end{split}
\end{align}
where $k_{1}$ and $k_{2}$ denote the momenta of the intermediate vector boson, 
$\gamma$ or $Z$ and outgoing vector boson $Z$ respectively and $a_V$, $b_V$ 
and $\tilde{b}_V$,  in general complex, are form factors.

Here we consider in a general model-independent way the production of a Higgs 
mass eigenstate $H$ through the process $e^+e^- \to HZ$ mediated by $s$-channel 
virtual $\gamma$ and $Z$. We have studied how beam polarization 
(both transverse and longitudinal) can be used to constrain anomalous 
interactions of the Higgs with neutral electroweak bosons. There have been various recent works 
on the study of anomalous $VVH$ couplings at a linear collider \cite{han,biswal,Hag,dutta}. 
However, our approach is different from theirs in that we include $\gamma ZH$ 
coupling, study the effects of both longitudinal as well as transverse $e^+$ and $e^-$ beams and 
construct simple asymmetries and correlations based 
on the $CP$ and $T$ transformation properties \cite{SDR}. 
\vspace{-0.3cm}
\section{Use of beam polarization}\vspace{-0.3cm}
Polarized beams are likely to be available at ILC, and several studies have shown the 
importance of longitudinal polarization in reducing backgrounds and improving the sensitivity 
of new effects. With longitudinal polarization, the cross section for $e^+e^-$ annihilation is given by
% \bea
$$\sigma =(1-\pl\plbar) \left[\sigma_0+\peff \frac{\sigma_{RL}-\sigma_{LR}}{4}\right],$$
%   &=& 2(\mathcal L_{eff}/\mathcal L)[\sigma_0-{\red {\peff}} \frac{\sigma_{RL}-\sigma_{LR}}{4}]
% \eea
 where $\peff=\frac{\pl-\plbar}{1-\pl\plbar}$ is the ``effective polarization'', $\sigma_0$ is 
the unpolarized cross section, $\sigma_{LR}$ is the cross section with left helicity $e^-$ and right
 helicty $e^+$, and $\pl$ and $\plbar$ are degrees of longitudinal polarization 
of the $e^-$ and $e^+$ beams respectively.
 With opposite signs of beam polarization for $e^+$ and $e^-$, 
the factors $\peff$ and ($1-\pl\plbar$) are large and hence the corresponding 
total cross section is enhanced.
Since polarized beams may not be available for the full
period  of operation of the collider,
we consider alternative options of integrated luminosities for individual
combinations of polarization. We 
assume three runs of the collider with 
three different combinations:
a) unpolarized beams, b) same signs of beam polarization, and c) 
opposite signs of beam polarization for $e^-$ and $e^+$. Half of the integrated 
luminosity is assumed for the run a) with unpolarized beams and other half to be
 equally divided between the other two runs b) and c).

% \begin{center}
%  \begin{tabular}{|ll||l|l|}
% \hline
% & & $P_L^{eff}$ & $\mathcal L_{eff}/\mathcal L$\\\hline
% $\pl=0$, & $\plbar=0$ & $0\%$ & $0.50$\\
% $\pl=80\%$, & $\plbar=0$ & $80\%$ & $0.50$\\
% $\pl=-80\%$, & $\plbar=60\%$ & $-95\%$ & $0.74$\\
% $\pl=80\%$, & $\plbar=60\%$ & $39\%$ & $0.26$\\\hline
% \end{tabular}
% \end{center}
\vspace{-0.3cm}
\section{Angular distribution}\vspace{-0.3cm}
We calculate the angular distribution arising from the square of the SM amplitude and 
from the interference between the SM amplitude and the amplitude arising from the 
anomalous $ZZH$ and $\gamma ZH$ couplings while ignoring terms bilinear in anomalous 
couplings assuming new physics contributions to be small. We have treated the 
two cases of longitudinal and transverse polarizations for the electron and 
positron beams separately. In all these cases, we have calculated traces using 
the symbolic manipulation program `\textbf{FORM}' \cite{form}.

\begin{center}
 \textbf{A. Angular distributions for longitudinally polarized beams}
\end{center}\vspace{-0.2cm}
The angular distribution for the process $e^{+}e^{-}\rightarrow ZH$ with
longitudinal beam polarization and including 
anomalous $ZZH$ and $\gamma ZH$ contributions
may be written as
\begin{equation*} 
  \frac{d\sigma_{Z,\gamma}}{d\Omega}\propto\left(1-P_L\overline{P}_L \right)
  \Big[A^{Z,\gamma}(1+\sin^2\theta)+B^{Z,\gamma}+C^{Z,\gamma}{\cos\theta} 
\Big]
\end{equation*}
%  and
% \begin{align}\label{lgzh}
%  \begin{split}
%   \frac{d\sigma_{\gamma}^L}{d\Omega}&\propto\left(1-P_L\overline{P}_L 
% \right)\Bigg[A_L^{\gamma}+B_L^{\gamma}\sin^2\theta+C_L^{\gamma}\cos\theta 
% \Bigg],
%  \end{split}
% \end{align}
% the anomalous $\gamma ZH$ contribution. 
Apart from kinematic factors, the dependence of the 
coefficients $A$'s, $B$'s, etc on anomalous couplings 
and vector and axial couplings $g_V$ and $g_A$ is shown in Table \ref{tab:coeff}.
  \begin{table}[!h]
\begin{center}
\begin{tabular}{|c|c|c|c|}\hline
  & $Z$ & $\gamma$&Couplings\\\hline
$A$ & $(g_V^{2}+g_A^{2}) -2g_V g_A  \peff$ 
& $g_V - g_A  \peff$ &$\azr,\agr$\\
$B$ & $(g_V^{2}+g_A^{2})-2g_V g_A  \peff$ 
& $g_V- g_A  \peff$ &$\bzr,\bgr$\\
$C$ & $(g_V^{2}+g_A^{2}) \peff-2g_V g_A $ 
& $g_A - g_V  \peff$ &$\bzti,\bgti$ \\\hline
 \end{tabular}\caption{\label{tab:coeff}Dependence of coefficients $A$'s,$B$'s 
etc on 
vector and axial vector couplings and the anomalous couplings associated with them.}
\end{center}
  \end{table}
%  \end{wraptable}  \\
\vspace{-0.5cm} 

Immediate inferences from these expressions and Table \ref{tab:coeff} are: \\
% \begin{wraptable}{l}{0.68\columnwidth}
(i) If coefficients $A$, $B$ and $C$ could be 
determined independently it would be possible to determine the anomalous couplings 
$\agr$, $\azr$, $\bgr$, $\bzr$, $\bgti$ and $\bzti$.\\ (ii) Couplings $\azr$, $\agi$,
 $\bzi$, $\bgi$, $\bztr$, $\bgtr$ do not contribute to 
the angular distribution at this order, and hence remain undetermined.\\
(iii) Numerically $g_V$ is small, while $g_A=-1$. Hence, in unpolarized case, 
the terms $A^Z$, $B^Z$ and $C^{\gamma}$ dominate.
%  If these coefficients are 
% determined from angular distributions, it would be possible to measure 
% $\Re{\Delta a_Z}$, $\Re{b_{Z}}$ and $\Im \tilde b_\gamma$ almost independent 
% of one another.
 On the other hand, there would be very low sensitivity to the 
remaining couplings, viz., $\agr$, $\bgr$  and $\bgti$. \\
(iv) With longitudinal 
polarization turned on, with a reasonably large value of  $\peff$, the coefficients 
$C^Z$, $A^\gamma$ and $B^\gamma$ would become significant. In that case, 
the sensitivity to $\agr$, $\bgr$ and $\bzti$ would be significant.

\begin{center}
 \textbf{B. Angular distributions for transversely polarized beams}
\end{center}\vspace{-0.2cm}
The angular distributions for the process $e^{+}e^{-}\rightarrow ZH$ with
transverse beam polarization is 
% for the anomalous $ZZH$ and $\gamma ZH$ contributions is :
% \begin{align}\label{tsm}
%  \begin{split}
%   \frac{d\sigma_{SM}^T}{d\Omega}&\propto A_T^{SM}+B_T^{SM}\sin^2\theta
% +P_T\overline{P}_TC_T^{SM}\sin^2\theta\cos2\phi,
%  \end{split}
% \end{align}
%  the SM contribution,
\begin{align}\label{tvzh}
 \begin{split}
  \frac{d\sigma_{Z,\gamma}^T}{d\Omega}\propto 
% A_T^{Z,\gamma}+B_T^{Z,\gamma}\sin^2\theta+C_T^{Z,\gamma}\cos\theta 
\frac{d\sigma_{UP}^T}{d\Omega}
+P_T\overline{P}_T\sin^2\theta[D_T^{Z,\gamma}\cos2\phi+E_T^{\gamma}\sin2\phi], 
 \end{split}
\end{align}
  where, $d\sigma_{UP}^T/d\Omega$ is the differential cross section 
with unpolarized beams.

% \begin{align}\label{tgzh}
%  \begin{split}
%   \frac{d\sigma_{\gamma}^T}{d\Omega}&\propto      
% A_T^{\gamma}+B_T^{\gamma}\sin^2\theta+C_T^{\gamma}\cos\theta\\
% &+P_T\overline{P}_T\bigg\{   D_T^{\gamma}\sin^2\theta
% \cos(2\phi-\delta)\\
% &+E_T^{\gamma}\sin^2\theta\sin(2\phi-\delta)\bigg\},
%  \end{split}
% \end{align}
% the anomalous $\gamma ZH$ contribution.
From the above expression one can infer the following: \\
(i) To study any effects of transverse polarization, both electron 
and positron beams have to be polarized, \\
(ii) If the azimuthal angle $\phi$ of $Z$ is integrated over, 
there is no difference between the transversely polarized and unpolarized cross sections,\\
(iii) The $\cos2\phi$ term of angular distribution (the $D_T$ term) 
determines a combination only of the couplings $\azr$ and $\agr$,\\
(iv) A glaring advantage of using transverse polarization would be 
determination $\agi$ from the $\sin2\phi$ dependence of the angular 
distribution, \\
(v) Couplings $\azi$, $\bzi$, $\bgi$, $\bztr$, $\bgtr$ remain undetermined 
with either longitudinal or transverse polarization. 
\vspace{-0.3cm}
\section{ Observables}\vspace{-0.3cm}
We construct observables having definite $CP$ and $T$ transformation properties 
so as to probe couplings corresponding to interactions 
having those transformation properties under $CP$ and $T$. We have divided 
the observables into two categories : 
a) using the momenta of the $Z$ boson (listed in table \ref{obs-Z}), 
b) using the momenta of leptons coming from $Z$ decay (listed in table \ref{obs-lep}).
% In this section, we will discuss observables like forward-backward asymmetry, 
% and azimuthal asymmetries which can be used to determine anomalous couplings.
From Table \ref{obs-Z}, we can see that with observables which are constructed 
from $Z$ boson momenta, couplings $\azi$, $\bzi$, $\bgi$, $\bztr$, $\bgtr$ can not be probed. 
Since these couplings are $T$ odd, we need $T$-odd observables to probe them. 
 $T$-odd 
observables include vector triple products, and so we need additional vectors
which 
are only be available 
if we utilize the momenta of leptons coming from $Z$ decay and/or the spins of the initial 
$e^-$ and $e^+$. We have constructed such $T$ odd observables and have been listed in Table \ref{obs-lep}.
\begin{table}[ht]
\begin{center}\vspace{-0.5cm}
\begin{tabular}{cccccc}
\hline
\hline\\

Symbol & Observable & $CP$  & $T$ & Couplings\\ 
       &            &            &  		&  		 \\ 
\hline\\
 $A_{FB}$ 	& $sign[(p_1-p_2).q]$
  & $ - $ 	& $ + $ & $ \bzti,~\bgti $\\
%  $\frac{1}{\sigma}$ $[\sigma(\cos\theta\ge0)-\sigma(\cos\theta\le0)]$

 $A_T$	& $sign(\sin2\phi)$ 	
 & $ - $ 	& $ + $ & $ \agi$ \\
 
%  $A_T^{\prime}$ 	& $sign(\cos2\phi)$ 
%  & $ - $ 	      & $ - $ & $ \azr,~\agr $ \\

 $X_1$ 	& $(p_1-p_2).q$ 	
 & $ - $ 	& $ + $ & $ \bzti,~\bgti $ \\
  
 $X_7$ 	& $[(p_1-p_2).q]^2$ 	
  & $ + $ 	        & $ + $ & $ \bzr,~\bgr $\\

$Y_1$ 	& $(q_xq_y)$ 	
 & $ + $ 	& $ - $	 & $ \agi$ \\

 $Y_2$ 	& $(q_x^2-q_y^2)$ 	
 & $ + $ 	& $ + $	 & $ \bzr~,\agr~,\bgr $ \\
 \\\hline
\hline
\end{tabular} 
\caption{\label{obs-Z}Observables constructed using $Z$ boson momenta $q$, their CP and T properties and 
the couplings they probe. Here, $P=p_1+p_2$ and $q=p_{3}+p_{4}$}.
\end{center}\vspace{-0.5cm}
\end{table}

Observables $X_i$ are sensitive to longitudinal polarization and $Y_i$ are sensitive to 
transverse polarization.
\begin{table}[!h]
\begin{center}
\begin{tabular}{cccccc}
\hline
\hline\\

Symbol & Observable & $CP$  & $T$ & Couplings\\ 
       &            &            &  		&  		 \\ 
\hline\\
 $X_2$	& $P.(\plm-\plp)$ 	
 & $ - $ 	& $ + $ & $ \bzti,~\bgti $ \\
 
 $X_3$ 	& $(\overrightarrow{\plm}\times \overrightarrow{\plp})_z$ 
 & $ - $ 	      & $ - $ & $ \bztr,~\bgtr $ \\

 $X_4$ 	& $(p_1-p_2).(\plm-\plp)(\overrightarrow{p}_{l^-}\times \overrightarrow{p}_{l^+})_z$ 	
 & $ - $ 	& $ - $ & $ \bztr,~\bgtr $ \\
  
 $X_5$ 	& $(p_1-p_2).q(\overrightarrow{\plm}\times \overrightarrow{\plp})_z$ 	
 & $ + $ 	& $ - $ & $ \bzi,~\bgi$ \\
 
 $X_6$ 	& $P.(\plm-\plp)(\overrightarrow{\plm}\times \overrightarrow{\plp})_z$ 	
 & $ + $ 	& $ - $ & $ \bzi,~\bgi$ \\
 
 $X_8$ 	& $[(p_1-p_2).(\plm-\plp)]^2$ 	
  & $ + $ 	        & $ + $ & $ \bzr,~\bgr $\\

 $Y_3$ 	& $(\plm-\plp)_x(\plm-\plp)_y$ 	
 & $ + $ 	& 	$ - $	 & $\agi,~\bgi$\\
 
 $Y_4$ 	& $q_xq_y(\plm-\plp)_z$  	
 & $ - $ 	& 	$ - $	 & $ \bztr,~\bgtr$ \\
 
 $Y_5$ 	& $(\plm-\plp)_x(\plm-\plp)_y q_z$	
 & $ + $ 	& 	$ - $	 & $\bzi,~\bgi $\\
 
 $Y_6$ 	& $[(\plm)_x^2-(\plp)_x^2]-[(\plm)_y^2-(\plp)_y^2]$	
 & $ - $ 	& 	$ + $	 & $\bzti,~\bgti $\\
 \\\hline
\hline
\end{tabular} 
\caption{\label{obs-lep}Observables constructed using lepton momenta from $Z$ decay, their CP and T properties and 
the couplings they probe. Here, $P=p_1+p_2$ and $q=p_{3}+p_{4}$}.
\end{center} \vspace{-1.0cm}
\end{table}
\vspace{-0.3cm}
\section{Kinematical cuts}\vspace{-0.3cm}
We need the following kinematical
cuts for the identification of the decay leptons :\\
% \vskip 0.1cm
 1. $E_f\geq10$ GeV for each outgoing charged lepton,	\\
 2. $5^{\circ}\leq \theta_0\leq 175^{\circ}$ for each outgoing
charged lepton for it to remain away from the beam pipe,\\
 3. $\Delta R_{ll}\geq0.2$ for the pair of charged lepton, where 
       $(\Delta R)^2\equiv(\Delta \phi)^2+(\Delta \eta)^2$, $\Delta
\phi$ and $\Delta \eta$ being the separation in azimuthal angle and rapidity,
respectively\\
 4. Cut on the invariant mass of the
$f\overline{f}$ to confirm  onshellness of the $Z$  boson, which is
$R1\equiv |m_{f\overline{f}}-M_Z|\leq 5\Gamma_Z$. 
This cut also reduce the contamination from  $\gamma\gamma H$ couplings.
\vspace{-0.3cm}
\section{Sensitivities}\vspace{-0.3cm}
We now obtain numerical results for the asymmetries, the correlations and 
the sensitivities of these observables for a definite configuration of the 
linear collider. For our numerical calculations, we have made use of the 
following values of parameters: $M_Z=91.19$ GeV, $\alpha = 1/128$, 
$\sin^2\theta_W = 0.22$. For the parameters of the linear collider, 
we have assumed $\sqrt{s} = 500$ GeV, $P_L = 0.8$, $\overline{P}_L = \pm 0.6$, 
$P_T = 0.8$, $\overline{P}_T = \pm 0.6$, and an integrated luminosity 
$\int \mathcal L dt = 500$ fb$^{-1}$. We have chosen to work with Higgs mass of $120$ GeV.

Each observable depends
on a combination of a limited number of couplings, dependent on CP and T
properties. We can determine, from a single observable, 
limits on individual couplings by combining the results from more than one observable,
or from more than one combination of polarization.
We will refer to limits on a coupling as an 
individual limit if all other couplings are zero.
If no such assumption is made, and more than one observable is used
simultaneously to put limits on all couplings contributing to these
observables, we will refer to the limits as simultaneous limits.

We demand that the contributions to the observable coming from the
anomalous parts are less than the statistical fluctuation in these
quantities at a chosen level of significance and study the sensitivity
of a LC to probe them. 

Statistical fluctuations in the cross-section or in an asymmetry
for a given luminosity ${\cal L}$ can be written as:
% \begin{eqnarray}
$\Delta\sigma = {\sqrt{\sigma_{SM}/{\cal L} }}
~~{\rm and}~~ (\Delta A)^2 = (1-A^2_{SM})/(\sigma_{SM}{\cal L})$ respectively.
% \end{eqnarray}

Transverse polarization, using $A_T$, enables
the determination of $\agi$ independent of all other couplings, with a
possible 95\% CL limit of about 0.04.

For the purpose of illustrating how we determine simultaneous limits, let us 
 take, \begin{wrapfigure}{r}{0.55\columnwidth}\vspace{-0.4cm}
\begin{center}
 \includegraphics[scale=0.55]{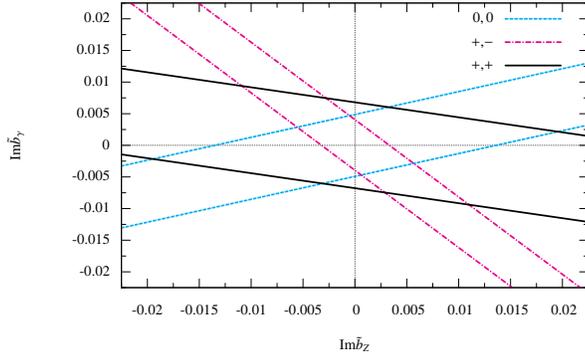}\vspace{-0.2cm}
 % fig1.eps: 0x0 pixel, 300dpi, 0.00x0.00 cm, bb=143 323 542 562
\caption{\label{sim:Afb}The region in the $\bzti$-$\bgti$ plane accessible at the $95$\% CL with observable 
$X_1$ with different longitudinal beam polarization configurations. 
$(0,0)$, $(+,-)$ and $(-,+)$ stand for 
$(P_L,\overline{P}_L)=$ $(0,0)$, $(0.8,-0.6)$ and $(-0.8,0.6)$ respectively.}
\end{center}\vspace{-0.8cm}
\end{wrapfigure}
for example, the observable $A_{FB}$ which 
determines a combination of 
$\bzti$ and $\bgti$. For three different polarization combinations chosen by us,
 we have three different combinations of these couplings determined by $A_{FB}$ 
which we have plotted in fig \ref{sim:Afb}. The best simultaneous limits is 
obtained by looking at the smallest region enclosed by the lines. 
In fig \ref{sim:Afb}, the smallest region is enclosed by lines corresponding to 
unpolarized beams and beams having opposite sign of polarization of $e^+$ and $e^-$. 
The best limits are obtained by looking at the extremeties of the region.  
These are $\vert \mathrm Im \tilde{b}_{\gamma}\vert \leq 4.69\cdot 10^{-3};  
\vert \mathrm Im \tilde{b}_Z\vert \leq 5.61 \cdot 10^{-3}$. 
In a similar fashion, we can put simultaneous limits on all other couplings by using 
different observables or using various combination of polarizations. 
In Table \ref{afblim}, we show $95$\% individual limits obtained on couplings 
$\bzti$ and $\bgti$ using the forward-backward asymmetry $A_{FB}$ for various 
combinations of beam polarizations.
\begin{table}[!h]
\begin{center}
\begin{tabular}{|l|c|c|}
\hline
&$\vert \bgti\vert$ & $\vert \bzti\vert$
\\ \hline
Unpolarized &
 0.00392 & 0.0108 \\
$P_L=0.8,\,\overline P_L=+0.6$ &
0.00543  & 0.0229\\
$P_L=0.8,\,\overline P_L=-0.6$ &
0.00320  & 0.00262 \\
\hline
\end{tabular}\caption{\label{afblim}95 \% CL individual limits on 
couplings $\bzti$ and $\bgti$ using forward-backward asymmetry, $A_{FB}$.}
\end{center}\vspace{-0.8cm}
\end{table} 

In Tables \ref{translim500} and \ref{longlim500}, we show $95$\% CL individual limits on 
anomalous couplings using observables for different combinations of beam polarizations.
From Table \ref{longlim500}, we see that longitudinal polarization helps in enhancing 
the sensitivities of the couplings which are not much sensitive in unpolarized case.

\begin{table}[!h]
\begin{center}
\begin{tabular}{cccc}
\hline
\hline
\multicolumn{3}{c}{} & $(P_T,\overline{P}_T)$
\\
\multicolumn{2}{c}{Observable} & Coupling   & $(0.8$,\,  $\pm 0.6)$\\ 
\hline
  $Y_1$         & $(q_x q_y)$   
 & $\agi $              & $1.98\times 10^{-1}
$\\

 $Y_2$  & $(q_x^2-q_y^2)$       
%  & $ \agr $       & $8.15\times 10^{-1} $\\
 & $ \bzr $       & $2.65\times 10^{-2} $\\
%  && $ \bgr $       & $3.41\times 10^{-1} $\\

 $Y_3$  & $(\plm-\plp)_x(\plm-\plp)_y$  
%  & $\agi $                      & $9.62 $\\
  & $\bgi $                    & $4.72\times
10^{-2}$ \\

 $Y_4$  & $q_xq_y(p_{l^-}-p_{l^+})_z$   
 & $\mathrm Im b_Z $             & $1.58\times 10^{-1}$\\
%  & & $\mathrm Im b_{\gamma} $     & $1.96 $ \\

  $Y_5$         & $(\plm-\plp)_x(\plm-\plp)_y q_z$      
  & $\bztr $                     & $5.56\times
  10^{-2} $\\
%  & & $\bgtr $                   & $6.89\times
%  10^{-1}$ \\

  $Y_6$   &  $ [(\plm)_x^2-(\plp)_x^2] -[(\plm)_y^2-(\plp)_y^2]$
  & $\bzti$                               & $1.10\times10^{-1}$\\
%  & %$-[(p_{l^-})_y^2-(p_{l^+})_y^2] $
%  & $\bgti$ 				 & $1.36 $\\

 \hline
\hline
\end{tabular} 
 \caption{\label{translim500}The $95$ \% C.L. individual limits on the anomalous
 $ZZH$ and $\gamma ZH$ couplings, with transversely polarized beams 
 for $\sqrt{s}=500$ GeV and $\int \mathcal L ~dt=500$ fb$^{-1}$. }
\end{center}
\end{table}

\begin{table}[!h]
\centering
\begin{tabular}{cccccc}
\hline
\hline
\multicolumn{2}{c}{} & Coupling  & $P_L=0$ & $P_L=0.8$ & $P_L=0.8$\\ 
       &            &            & $\overline{P}_L=0$ &
$\overline{P}_L=0.6$ & $\overline{P}_L=-0.6$\\ 
\hline
 $X_1$  &           
  & $\bzti $    & $4.11\times 10^{-2} $ & $8.69\times 10^{-2} $&
$9.94\times 10^{-3}$\\
  & & $\bgti $  & $1.49\times 10^{-2} $ &$2.06\times 10^{-2} $
&$1.22\times 10^{-2} $ \\

 $X_2$       & 
 & $\bzti $     & $4.12\times 10^{-2} $ & $5.99\times 10^{-2} $ &
$3.84\times 10^{-2} $\\
 & & $\bgti $   & $5.23\times 10^{-1} $ & $3.12\times 10^{-1} $&
$5.52\times 10^{-2} $\\

 $X_3$  & 
 & $\bztr $           & $1.41\times 10^{-1} $ & $2.97\times 10^{-1} $ &
$3.40\times 10^{-2} $\\
 & & $\bgtr $ & $5.09\times 10^{-2} $ & $7.05\times 10^{-2} $ &
$4.15\times 10^{-2} $ \\

 $X_4$  & 
 & $\bztr $     & $2.95\times 10^{-2} $ & $4.29\times 10^{-2} $ &
$2.75\times 10^{-2} $\\
&  
  & $\bgtr $   & $3.81\times 10^{-1} $ &$ 2.24\times 10^{-1}$ &
$3.95\times 10^{-2}$ \\

  $X_5$  & 
  & $\bzi $      & $ 7.12\times 10^{-2}$ & $ 1.04\times 10^{-1}$ & $
 6.64\times 10^{-2}$\\
  & & $\bgi $    & $ 9.10\times 10^{-1}$ & $ 5.42\times 10^{-1}$ &
 $9.53\times 10^{-2}$ \\

  $X_6$  & 
  & $\bzi $      & $ 7.12\times 10^{-2}$ & $ 1.50\times 10^{-1}$ & $
 1.72\times 10^{-2}$\\
  &       
 & $\bgi $ & $ 2.58\times 10^{-2}$ & $ 3.57\times 10^{-2}$ &
 $2.10\times 10^{-2}$ \\

  $X_7$  & 
  & $\bzr $      & $1.75\times 10^{-2} $ & $2.54\times 10^{-2} $&
 $1.63\times 10^{-2} $\\
  &
  & $\bgr $& $2.23\times 10^{-1} $ &$ 1.34\times 10^{-1}$ &
 $2.35\times 10^{-2}$ \\ 

  $X_8$       &
  & $\bzr $      & $1.53\times 10^{-2} $ & $2.22\times 10^{-2} $&
 $1.42\times 10^{-2} $\\
  & 
 & $\bgr $    & $1.94\times 10^{-1} $ &$ 1.16\times 10^{-1}$ &
 $2.04\times 10^{-2}$ \\ 

\hline
\hline
\end{tabular} 
\caption{\label{longlim500} The $95$ \% C.L. individual limits on the anomalous
$ZZH$ and $\gamma ZH$ couplings, for $\sqrt{s}=500$ GeV and integrated luminosity $\int \mathcal L
~dt=500$ fb$^{-1}$.}\vspace{-0.5cm}
\end{table}
\vspace{-0.3cm}
\section{Summary and conclusions}\vspace{-0.3cm}
We have obtained angular distributions for the process  $e^+e^-\rightarrow ZH$ in the presence of
anomalous $\gamma ZH$ and $ZZH$ couplings to linear order in these couplings in the presence of
longitudinal and transverse beam polarizations. We have then looked at observables and
asymmetries which can be used in combinations to disentangle the various couplings to the
extent possible. We have also obtained the sensitivities of these observables and asymmetries
to the various couplings for a definite configuration of the linear collider.

In some cases, the contribution of some couplings is suppressed due to the vector coupling of 
$Z$ to $e^+e^-$ being very small. In those cases longitudinal polarization helps to enhance 
the contribution of these couplings and thereby improving the sensitivity.

We find that with a linear collider operating at a c.m. energy of 500
GeV with the capability of 80\% electron polarization and 60\% positron
polarization with an integrated luminosity of 500 fb$^{-1}$, 
with the observables described above, it would be
possible to place 95\% CL individual limits of the order of a few times 
$10^{-2}$ on all couplings taken 
nonzero one at a time with use of an appropriate combination ($P_L$ and
$\overline P_L$ of opposite signs) of longitudinal beam polarizations.
This is an improvement by a factor of 5 to 10 as compared to the
unpolarized case. 
The simultaneous limits possible are, as expected, less stringent. While they
continue to be better than $5\times10^{-2}$ for most couplings.
Transverse polarization enables
the determination of $\agi$ independent of all other couplings.
% Also, transverse polarization helps to probe a combination of $\azr$ and $\agr$ 
% independently of $\bzr$ and $\bgr$. 
Also, in the angular distribution of $Z$, all $T$ odd 
anomalous couplings $\azi$, $\bzi$, $\bgi$, $\bztr$ and $\bgtr$ are absent, and 
therefore, inorder to probe these couplings, we need to construct $T$-odd observables which involve vector triple products. 
For this, we study the decay of the $Z$ boson to a charged lepton pair and 
construct various $T$-odd observables using the momenta of the leptons coming from $Z$ decay. With $T$-odd 
observables, we could probe all the anomalous couplings which are earlier absent.

We have assumed that only one leptonic decay mode of $Z$ is observed.
Including both $\mu^+\mu^-$ and $\tau^+\tau^-$ modes would trivially
improve the sensitivity. In case of observables like $X_1$, $Y_1$, $Y_2$,
which do not need charge identification, even hadronic decay modes of
$Z$ can be included, which would considerably enhance the sensitivity.
\begin{footnotesize}

\end{footnotesize}
\end{document}